\documentclass{article}

\begin{document}

{\Large{\centerline{\bf  Entropy of Static Spacetimes}}}
{\Large{\centerline{\bf and Microscopic Density of States}}}

\bigskip

\centerline{T.~Padmanabhan}

\medskip
\centerline{IUCAA, 
Post Bag 4, Ganeshkhind, Pune - 411 007, India}
\centerline{Email: nabhan@iucaa.ernet.in}


\bigskip

\begin{abstract}

A general ansatz for gravitational entropy can be provided using the criterion that, any patch of area
which acts as a horizon for a suitably defined accelerated observer, must have an entropy proportional to its area. After providing a brief justification for this ansatz, several consequences are derived: (i) In any static spacetime with a horizon and associated temperature $\beta^{-1}$, this entropy satisfies the relation
$S=(1/2)\beta E$ where  $E$ is the energy source for gravitational acceleration, obtained as an integral of
$(T_{ab}-(1/2)Tg_{ab})u^au^b$. (ii) With this ansatz of $S$, the minimization of Einstein-Hilbert action is equivalent to minimizing the free energy $F$ with $\beta F=\beta U-S$ where $U$ is the integral of $T_{ab}u^au^b$. We discuss the 
conditions under which these results imply $S\propto E^2$ and/or $S\propto U^2$ thereby generalizing
the results known for black holes.
 This approach links with several other known results, especially the
holographic views of spacetime.
 
\end{abstract}

\bigskip

\section{Introduction and Motivation}

It has been known for a long time that spacetimes with horizons show an intriguing similarity with thermodynamic
systems. (For a review and detailed references to previous work, see the standard text book \cite{birrel}
or the recent review \cite{tprmp}).  A wide class of static
or stationary horizons studied in the literature leads naturally to a periodicity criterion in the Euclidean time coordinate, which can be used to motivate the association of a temperature with the horizon. In the familiar situations,
the temperature $T$ is related to the magnitude of the surface gravity $\kappa$ of the horizon by $T=\kappa/2\pi$.
This result has been derived from many different perspectives and the connection between horizon and temperature seems to be non controversial.

It is, of course, natural to ask whether one can associate other thermodynamical variables, like entropy, free energy etc. with spacetimes containing a static horizon. Since horizons block information to class of observers, it seems reasonable to associate an entropy with the horizons. (In fact, historically, Bekenstein attributed \cite{beken} entropy
to black hole horizon {\it before} temperature was attributed to it !).  There seems to be  less unanimity among
researchers whether \emph{all} horizons have entropy.  Part of the problem arises because --- in general --- horizons 
can be observer dependent while  one is tempted to think of entropy as an observer independent attribute. And, of course, unless one can attribute entropy to horizons, one cannot proceed further to other thermodynamical variables.

In this paper, we shall make an attempt to attribute an entropy to any horizon by providing a suitable definition and investigating the consequences. (Here, and in what follows, we use the term `definition of entropy' to mean a suitable mathematical
expression for entropy in a specific context. Of course, in standard thermodynamics or statistical mechanics, entropy is already `defined' and in that sense, we are providing an ansatz rather than a definition. We will not make this semantic distinction, since it will be clear from the context what we mean by `definition'.) There are two key ingredients which go into this analysis: 

First, we take the point of view that if
 any  family of observers do not have access to part of the spacetime, then they will attribute an entropy $S$ to the gravitational field because of the  degrees of freedom
which are hidden. Such families of observers exist in any spacetime and can be characterized as follows: Consider a congruence of time like curves in a region of spacetime. If the union of the causal past of this congruence has a nontrivial boundary, that boundary will act as a horizon for these observers.  Familiar examples are uniformly accelerated observers in Minkowski spacetime or observers following $r=$ constant ($>2M$) world lines in the Kruskal
spacetime etc. An observer who crosses over to $r<2M$ region of Kruskal spacetime can certainly access more
information than the observer who stays outside the horizon forever; if entropy of the black hole is related to lack of information, it is clear that it depends on which class of observers one is considering, even in the case of Schwarzschild horizon which can be defined in a purely geometric manner. We believe that, just as one had to give up the notion of particles being observer independent construct, we also need to give up the notion of entropy being an observer independent construct. (The idea that {\it all} horizons have entropy has been discussed in detail in \cite{tprmp}
and in \cite{reftwo}).

Second, even if entropy is observer dependent,  it has to be generally covariant. That is, we are free to use the variables (like the four velocity $u^a$) associated with a given congruence of observers to define the entropy but the definition should be a generally covariant scalar. This scalar will depend on the variables associated with the congruence of observers and hence will change when the congruence is changed.
(This makes sense because different class of observers can access different amounts of information.)
 We shall show that it is indeed possible to provide such s definition which has several desirable properties. In particular, the same congruence of
 observers will  associate a temperature $\beta^{-1}$ with the horizon and will measure the  source of gravitational acceleration  $E$ and  the total energy $U$ in any region. (These differ because pressure contributes to gravitational acceleration, as discussed below). Once a suitable definition for entropy $S$ can be provided in terms of geometrical variables, then one can investigate how $S$ is related $U,E$ and $\beta$. We shall provide one such definition 
and study the consequences. 

\section{The entropy of static horizons}

To do this, we need to set up a geometric framework which is adapted to a congruence of observers who sees a horizon. This concept, closely related to the idea of a local Rindler frame, was explored in \cite{tprmp} and, more recently, in \cite{tpmpla04}. (In the latter work, it was shown that the notion of entropy arises very naturally from the tunneling probability across the horizon.) It is this coordinate system which we will use to provide a suitable definition for entropy. We  begin with a description of a static spacetime and define the relevant thermodynamic variables in this frame.

 The metric of  a static spacetime can be
put in the form
$ds^2=-N^2dt^2+\gamma_{\mu\nu} dx^\mu dx^\nu$
where $N$ and $\gamma_{\mu\nu}$ are independent of $t$. (We use the signature -- + + +; Greek indices cover 1,2,3 and Latin indices cover 0--3). The comoving observers at $x^\mu=$ constant have the four velocity
$u_i=-N\delta^0_i$ and the four acceleration $a^i=(0,\partial^\mu N/N)$. If $N\to 0$ on a two-surface, with
$ N^2 a^2\equiv(\gamma_{\mu\nu}\partial^\mu N\partial^\nu N)$ finite, then this coordinate system has a horizon and 
one can associate a temperature with this horizon. While  the temperature is easily obtained by transforming to the Rindler coordinates near the horizon, it is clearer in the following approach:

 We note that static spacetimes  have a natural coordinate system in terms of the level surfaces of $N$
 (which correspond to the equipotential surfaces in the limit of Newtonian gravity). That is, we transform from the original space coordinates $x^\mu$ to the set $(N,y^A), A=2,3$ by treating $N$ as one of the coordinates. The $y^A$ denotes the two
transverse coordinates on $N=$ constant surface. (Capital Latin letters go over the coordinates 2,3 on the $t=$ constant, $N=$ constant surface). This
can be always done locally, but possibly not globally, because $N$ could be multiple valued etc. We, however,
 need this description only locally.  The metric now becomes
\begin{equation}
ds^2=-N^2dt^2+ \frac{dN^2}{(Na)^{2}}+
\sigma_{AB}(dy^A-\frac{a^A dN}{Na^2})(dy^B-\frac{a^BdN}{Na^2})
\label{iso}
\end{equation}
The original 7 degrees of freedom in $(N,\gamma_{\mu\nu})$ are now reduced to 6 degrees of freedom in $(a,a^A,
\sigma_{AB})$, because of our choice for $g_{00}$. This reduction  is similar to what happens in 
the synchronous coordinate system which makes
$N=1$ but loses the static nature \cite{ll}. In contrast, Eq.(\ref{iso}) which describes the spacetime in terms of
the magnitude  of acceleration $a$, the transverse components $a^A$ and the  metric $\sigma_{AB}$ on the two surface, provides a nicer description of geometry, maintaining the $t-$ independence. 
The $N$ is now merely a coordinate and the spacetime geometry is  described in terms of 
$(a,a^A,\sigma_{AB})$
all of which are in general functions of $(N,y^A)$. In well known, spherically symmetric spacetimes with horizon, we will have $a=a(N),a^A=0$ if we choose $y^A=(\theta,\phi)$. In general, important features of dynamics are encoded in $a(N,y^A)$. 
(More detailed discussion  of this coordinate system is given elsewhere \cite{tpmpla04}).

 Near the $N\to 0$ surface, $Na\to \kappa$, the surface gravity, and the $N-t$ part of the metric reduces to the Rindler form $ds^2=-N^2 dt^2 + dN^2/\kappa^2 + ...$. Regularity on the Euclidean sector now requires periodicity of Euclidean time with period $|\beta|=2\pi/\kappa$. This allows us to define the temperature in terms of 
 the derivative of $N$, {\it whenever
there is a horizon}. [The sign of $\beta$ depends on the sign of the derivative of $N$ near the horizon. For example, it is positive for the Schwarzschild black hole horizon while it is negative for de Sitter
 horizon. The surface gravity $\kappa$ is taken to be positive by definition, which accounts for the modulus
 sign in $|\beta|=2\pi/\kappa$. It turns out that the relevant signs for different thermodynamic variables work out correctly in both the cases $\beta>0$ and $\beta<0$. When it is not important, we will disregard the sign of $\beta$
 --- that is, pretend it is positive --- and consider it as inverse temperature.]
 
 Having defined the temperature, we want to next define other thermodynamic variables like energy, entropy etc. To
 do this, it will turn out to be convenient to consider a compact region in spacetime
defined as follows: the
  3-dimensional spatial region  is taken
to be some compact volume
${\cal V}$ with boundary $\partial{\cal V}$. 
(To discuss Rindler type horizon, we need to consider $\partial \mathcal{V}$ that is non compact. We shall
discuss this separately below and it does not create any special difficulties.)
As regards the time variable, we note that the static spacetimes are
time translation invariant. Thus integration over $dt$ in physical quantities --- like for example, action ---
can be taken over any interval and will diverge if integrated over $(-\infty,+\infty)$.  However, when the spacetime has a horizon, there is a natural interval in the {\it Euclidean time}, viz. $(0,\beta)$ and one can work everything out in the Euclidean sector, keeping a finite time interval.  (We take $\beta$ to be positive for these considerations
or more generally, use $|\beta|$.) In fact, in the Euclidean sector, the horizons can be mapped to singular point
and the thermodynamic variables can be related to topological invariants \cite{tprmp,tptopo,teit}.

But since the concepts of time like congruence, observers etc 
(which we will be extensively using)
are 
defined in the Minkowski spacetime, it will be convenient if we can identify the corresponding range of integration in the Minkowski time.  In the case of static horizons, the analytic continuation to Euclidean time $t_E$ is made by writing $\tau(\theta)=t\exp(i\theta)$ and varying $\theta$ from $0$ to $\pi/2$ giving $t=\tau(0)$ and $t_E = \tau(\pi/2)$.
 Hence the interval
$(0,\beta)$ in $t_E$ can be mapped to the same interval $(0,\beta)$ in $t$ as well and we can continue to work in the Minkowski spacetime.  The mapping of the intervals is obvious from the geometric concept of Wick {\it rotation} and the justification for keeping this range finite arises from the underlying Euclidean theory which
has been discussed in several works \cite{tprmp,tptopo,teit}. Except for ``borrowing" the range of integration,
we can work in Lorentzian sector of the theory itself which proves to be convenient for introducing time like
congruences, horizons, regions on two sides of horizons etc. all of which are less concrete in the in
Euclidean sector.

We are now ready to introduce our ansatz (or `definition') for entropy.
We  define the entropy associated with the horizon perceived by a congruence of observers to be:
\begin{equation}
S=\frac{1}{8\pi G}\int\sqrt{-g}d^4x\nabla_i a^i=\frac{\beta}{8\pi G}\int_{\partial\cal V}\sqrt{\sigma}d^2x(Nn_\mu a^\mu)
\label{defs}
\end{equation}
The second equality is obtained because, (i)
for static spacetimes the time integration reduces to multiplication by $\beta$ because of reasons explained above
and (ii) since only the spatial components of
$a^i$ are non zero, the divergence  becomes a three dimensional one over ${\cal V}$
which is converted to an integration over its boundary ${\partial\cal V}$. 
When the horizon arises as a limiting sequence of compact surfaces (like in the case of Schwarzschild metric)
we will take ${\partial\cal V}$ to be this limiting surface. In our coordinate system, this arises as a limit of
$N({\bf x})=$ constant surfaces in the limit of the constant tending to zero. We shall see below that, if the spacetime is empty ($T_{ab}=0$), then one could use any other compact surface enclosing the horizon. When the horizon is
not compact (like in the case of Rindler coordinates), we shall use the second expression in Eq. (\ref{defs}) to define the contribution from the surface
${\partial\cal V}$ to the entropy. Once again, this is the $N\to 0$ surface in our coordinate system.

We take this quantity to be the definition of gravitational entropy for any static spacetime with a horizon, based on the following considerations:

(a) If the boundary ${\partial\cal V}$  is a standard black hole horizon, 
$(Nn_\mu a^\mu)$ will tend to a constant surface gravity $\kappa$  and the using $\beta\kappa=2\pi$ we get $S={\cal A}/4G$
where ${\cal A}$ is the area of the horizon. 
(In case of non compact horizons we obtain the entropy per unit area to be $(4G)^{-1}$; see item (b) below.)
Thus, in the familiar cases, this does reduce to the standard expression
for entropy \cite{beken,bhentropy}. We will see later that, if the boundary ${\partial\cal V}$ is a compact surface enclosing a compact horizon ${\cal H}$
and if the
region between ${\partial\cal V}$ and ${\cal H}$ is empty, then again we get the entropy $S={\cal A}/4G$
because the flux through the two surfaces are the same when the in between region has $T_{ab}=0$.
The numerical factors and sign in the ansatz is chosen so that the standard results are recovered and that the entropy is positive.

(b) Similar considerations apply to each piece of {\it any} area element when it acts as a horizon for some Rindler observer. Results obtained in a previous work \cite{tps} 
 showed that the bulk action for gravity can be obtained from a surface term in the action, if we take the entropy
of any horizon to be proportional to its area with an elemental area $\sqrt{\sigma}d^2x$ contributing an entropy
$dS=(Nn_\mu a^\mu)\sqrt{\sigma}d^2x$. Our definition is the integral expression of the same. Further, it was shown in
ref. \cite{tpmpla04} that the tunneling probability across the horizon leads to exactly the same identification for the entropy.

(c) A family of observers, defined through a congruence of time like curves,
will perceive a horizon, whenever the boundary of the union of causal past of the congruence is nontrivial.
If the entropy is related to the unobserved degrees of freedom blocked off by the horizon, then it {\it must} depend on the congruence of time like curves but be coordinate independent. This criterion is satisfied by our definition; the $a^i$ is defined using the congruence of comoving observers in Eq.(\ref{iso}), who have a horizon at $N=0$. (We do not provide detailed justification for this congruence dependent
entropy in this paper, since they are given in previous work \cite{tprmp,reftwo,tps}). 

(d) Finally, the definition shows that the entropy relevant for gravitational field in a bulk spatial region ${\cal V}$ resides on the  boundary $\partial{\cal V}$ of the region.
Such a holographic view has been advocated recently in several papers \cite{bousso}.
 In fact, defining the analogue
of the ``tortoise coordinate'' $q$ by $dq=-dN/N(Na)$, we see that $N\approx\exp(-\kappa q)$ near the horizon, where $Na\approx \kappa.$ The space part of the metric in Eq.(\ref{iso}) becomes, near the horizon
$dl^2=N^2(dq^2+e^{2\kappa q}dL_\perp^2)$
which is conformal to the AdS metric. The horizon, on which the degrees of freedom contributing to entropy live,
becomes the $q\to\infty$ surface of the AdS space. (These results hold in any dimension). All these strengthen the idea of
Eq.(\ref{defs}) providing the definition for gravitational entropy. This entropy measures
the flux of gravitational field lines through the surface. 
The flux through any horizon is quantized in the semi-classical limit \cite{tpapoorva} and this result ties up with the fundamental limitation in measuring intervals smaller than Planck scale \cite{tplp} and with quantization
of horizon areas \cite{beken,areaquant}. 

There is simple relation between the integrand in Eq.(\ref{defs}) and the matter energy momentum tensor present in the spacetime. In any  space time, there is differential geometric identity (see e.g. Eq. (B1) of \cite{tpapoorva})
\begin{equation}
R_{bd} u^b u^d 
= \nabla_i(Ku^i +a^i) - K_{ab}K^{ab} + K_a^a K^b_b
\label{idone}
\end{equation}
where $K_{ab}$ is the extrinsic curvature of spatial hypersurfaces and $K$ is its trace. This
reduces to
$\nabla_i a^i=R_{ab}u^au^b$ in static spacetimes with $K_{ab}=0$. Combined with Einstein's equations, this gives
\begin{equation}
\frac{1}{8\pi G}\nabla_i a^i= (T_{ab}-\frac{1}{2}Tg_{ab})u^au^b
\label{poisson}
\end{equation}
This equation deals directly with $a^i$ which occur as the  components of the metric tensor in
(\ref{iso}) and relates the integrand of Eq.~(\ref{defs}) to the matter stress tensor. 

Having introduced the concepts of temperature and entropy for static spacetimes with horizon, we shall next
turn to the concept of energy.  The most natural definition of energy in a region of space is given by the integral
\begin{equation}
U\equiv\int_{\cal V}d^3x\sqrt{\gamma}N(T_{ab}u^au^b)
\label{defofu}
\end{equation}
based on the energy density $\rho=T_{ab}u^au^b$.  An important consistency check on our definition of entropy
is the following:
If these ideas are consistent, then the free energy of the spacetime must have direct geometrical
meaning {\it independent of the congruence of observers used to define the entropy $S$ and $U$}.
The free
energy is defined as $F\equiv U-TS$ and is given by
\begin{equation}
\beta F\equiv \beta U -S =-S +\beta\int_{\cal V}d^3x\sqrt{\gamma}N(T_{ab}u^au^b)
\end{equation}
and using Eqs.(\ref{defs}),(\ref{poisson}) and $R=-8\pi G T$, we find that
\begin{equation}
\beta F=\frac{1}{16\pi G}\int d^4x\sqrt{-g}R
\label{ehfree}
\end{equation}
which is just the Einstein-Hilbert action. The equations of motion obtained by minimizing the action can be equivalently thought of as minimizing the macroscopic free energy. For this purpose, it {\it is} important that $F$ is generally covariant and is independent of the $u^i$ used in defining other quantities.

There have been several attempts in the past to obtain the Einstein's equations as a consequence of thermodynamics of the source under special assumptions and we shall briefly comment on them. The basic inspiration for this approach arises from the original work by
Sakharov \cite{sakharov} who interpreted gravity as due to elasticity of spacetime. One possible way of doing this
at the level of equations of motion was attempted in ref. \cite{ted}. However since the whole motivation is quantum mechanical (e.g. the entropy of a classical black hole is infinite), the action principle was better suited for this approach. This was definitely indicated in the work of Gibbons and Hawking \cite{gh} in which the Euclidean action was related to thermodynamic variables but no attempt was made to derive Einstein's equations as a thermodynamic identity.
This idea was explored in a more recent series of papers \cite{tprmp, tps} and an explicit demonstration that
Einstein's equations can be expressed as a thermodynamic identity was given, for spherically symmetric spacetimes, in ref. \cite{tpcqg}. The current work does not attempt to derive Einstein's equations from thermodynamic relation;
instead, it shows that there is a close relationship between thermodynamic variables and geometric variables
(entropy density and acceleration, free energy density and scalar curvature etc) when equations of motion are satisfied. In that sense, this work is more similar to that of ref. \cite{gh}. (For some other closely related work, see
ref. \cite{prev}).

While $U$ is the integral of the energy density as measured by the congruence of observers, it is {\it not} the source of gravitational acceleration. The latter has contribution from pressure as well  and is given by the 
Tolman-Komar energy  \cite{tolkomar} which is the source 
 for gravitational acceleration:
\begin{equation}
\label{defe}
E=2\int_{\cal V}d^3x\sqrt{\gamma}N (T_{ab}-\frac{1}{2}Tg_{ab})u^au^b
\end{equation}
The covariant combination  $2(T_{ab}-(1/2)Tg_{ab})u^au^b$ [which reduces to $(\rho+3p)$ for an ideal fluid]
 {\it is} the correct source for gravitational
{\it acceleration}. For example, this will make geodesics accelerate away from each other in a universe dominated by cosmological
constant, since $(\rho+3p)<0$.
The factor $N$ correctly accounts for the relative redshift of energy in curved spacetime. Since our metric components are now described, partly, by the  acceleration, the determination of metric is equivalent to determining the components of acceleration; the spatial part $\sigma_{AB}$ often plays a less important role. 
Integrating Eq.(\ref{poisson}) with the measure $\sqrt{-g}d^4x$ over a four dimensional region chosen
as before, and 
 using Eq.(\ref{defs}), Eq.(\ref{defe}), the integrated form of Eq.(\ref{poisson}) will read quite simply as
\begin{equation}
S=(1/2)\beta E,
\label{sofe}
\end{equation}

The sign of $E$ can be negative if matter with $\rho+3p<0$ dominates in the region ${\cal V}$. The sign of $S$
in (\ref{defs}) depends on the convention chosen for the direction of the normal to $\partial{\cal V}$
but it is preferable to choose this such that $S>0$. Then the sign of $\beta$ will arrange itself so that (\ref{sofe}) holds.
(Of course, the temperature is $T=|\beta|^{-1}>0$). 
As an illustration, consider the Schwarzschild spacetime and the de Sitter universe. For spherically symmetric metrics with a horizon, having  $g_{00}=-g^{11},g_{00}(r_H)=0$, we can write
$g_{00}\approx g'_{00}(r_H)(r-r_H)$ near the horizon and $\beta=-4\pi/g'_{00}(r_H)$ in our signature convention. Hence
$\beta=8\pi M>0$ for Schwarzschild while $\beta=-2\pi/H<0$ for de Sitter.
In the first case,
$\beta=8\pi M$ and we can take $E=M$ for any compact two surface $\partial{\cal V}$
that encloses the horizon. Since $Na=(M/r^2)$, Eq.(\ref{defs}) gives $S=4\pi(M^2/G)$ for any $\partial{\cal V}$.
This result agrees
with Eq.(\ref{sofe}). The de Sitter case is more interesting since it is nonempty. In the static coordinates with $-g_{00}=g^{rr}=(1-H^2r^2)$, let us choose a spherical surface of radius $L<H$. We have $E=-H^2L^3$ and $S=\pi H L^3$ from (\ref{defe}) and (\ref{defs}).
Once again, equation (\ref{sofe}) holds since $\beta=-2\pi/H$. Hence, as we said before, our definition ensures
correct signs for all physical variables.

\section{Thermodynamic interrelationships}

We shall next consider the question of relating $S$ directly to $E$ when $\partial{\cal V}$ is a horizon. {\it If}
 we can write  $\beta=(dS/dE)$, then Eq.(\ref{sofe}) can be immediately integrated to give 
\begin{equation}
S\propto E^2=(E/E_0)^2
\label{denstate}
\end{equation}
with some constant $E_0$, which can be determined  for any specific case in which the area of the horizon can be related to energy $E$, by demanding consistency of Eq.(\ref{denstate}) with $S={\cal A}/4G$, obtainable directly from Eq.(\ref{defs}). While the coefficient relating $S$ to horizon area is universal
and (1/4), the relation between $E$ and the horizon area depends on the spacetime. Hence $E_0$ has no universal expression.
This is understandable from the fact that entropy arising from the information hidden by the horizon is proportional to the
area of the horizon while the microscopic density of states will depend on the specific nature of the classical solution.
What is interesting is that the proportionality $S\propto E^2$ is universal whenever one can write $\beta=dS/dE$.
(A different approach to obtain $S\propto E^2$ and the area scaling of entropy was made in \cite{oppen} based on Gibbs relation).
The relation $T\propto E^{-1}$
 implies negative specific
heat and in-equivalence of different ensembles; this is in fact expected since they arise even in statistical mechanics of Newtonian gravitating systems (see e.g.,\cite{dlb}). 

This interpretation of $\beta=dS/dE$ is, however, nontrivial and this relation is {\it not} always valid with
{\it total} derivatives.. 
Usually in thermodynamics, one has the relation $\beta=(\partial S/\partial E)$ with {\it a partial derivative which requires other variables to be held fixed.}  As long as we consider the metric to be of the form $g_{ab}(x^\mu,p_i)$ where $p_i$'s are a set of parameters one can think of different ensembles of spacetimes in which different sets of parameters are held fixed. If all but one parameter is held constant (or in the simple cases in which the metric depends {\it only} on one variable like in the Schwarzschild metric or pure de Sitter) then one can relate $E,S$ etc to the single variable and obtain the result $S\propto E^2.$
But for a {\it general} solution, one cannot interpret
$dS/dE$ in any simple manner and hence one cannot proceed from (\ref{sofe}) to (\ref{denstate}). 
This is most easily
illustrated by considering a specific example rather than by abstract reasoning: Consider a class of spherically symmetric metrics with $-g_{00}=g^{rr}=f(r)=1-2m(r)/r$. This metric solves the Einstein's equations if the energy density $\rho(r)/8\pi$ and the transverse pressure $\mu(r)/8\pi$ are arranged to give $\rho(r)=(m'/2r^2); \mu(r)=\rho+(1/2)r\rho'(r)$ and the radial pressure is set equal to the energy density. If there is a horizon at $r=a$, with 
$f(a)=0,f'(a)\equiv B,\beta=4\pi/B$ then we will find that for a spherical region of radius $r=a$
\begin{equation}
S=\pi a^2; E=(1/2)a^2 B(a), B=(1/a)(1-4 a^2\rho(a))
\end{equation}
These relations hold on the horizon for a class of solutions parametrized by the function $m(r)$ with $a$
determined as the root of the equation $2m(a)=a$.
These $S,E$ and $\beta$ satisfy equation (\ref{sofe}) identically. But since
\begin{equation}
\frac{S}{E^2}=4\pi [ 1-2m'(a)]^{-2}
\end{equation}
equation (\ref{denstate}) is not obtained unless $m'(a)$ is a constant independent of the parameters of the solution.
This happens,  for example, in pure Schwarzschild spacetime with mass $M$ as well as pure de Sitter with Hubble parameter $H$. But in SdS, we get $S/E^2=\pi[(M/a)-1]^{-2}$ which is, in general, not a constant.
( The fact that SdS spacetime has multiple horizons does not affect the above equations which hold at each horizon separately.) From $S=\pi a^2, E=(1/2)a^2 B(a)$, it follows that 
\begin{equation}
\frac{dS}{dE}=\frac{\beta}{2}[1+(a/2B)(dB/da)]^{-1}
\end{equation}
which will not give $dS/dE=\beta$ except for the family of models for which $B\propto(1/a)$. (Note that, in the above equation $dS/dE$ is indeed the total derivative since we are now considering a family of solutions to  Einstein's equations parametrized by the horizon radius $a$.)
 
Remarkably enough, in these spacetimes,
 $U=a/2$ and hence $S\propto U^2$.
In fact, the relations $U=a/2,S=\pi a^2,\beta=4\pi/B$ along with the fact that radial pressure is equal
to the energy density allows us to  write Einstein's' equations as $dU=TdS-PdV$ 
(where the differentials are interpreted as $dU=(dU/da)da$ etc.). This was derived in ref. \cite{tpcqg} 
for the spherically symmetric spacetimes and it is not clear how to generalize it for non spherical cases. 
We suspect that the relation $S\propto U^2$ is more general (than, for example, $S\propto E^2$) and might hold even when the assumption of spherical symmetry is relaxed; but this issue requires further investigation.

A similar situation arises in 
Rindler spacetime with the metric in (\ref{iso}) where $a=\kappa/N,a^A=0$
and $\kappa$ is an arbitrary constant. The periodicity in Euclidean time will now give $\beta=2\pi/\kappa$.
The contribution to $S$ in (\ref{defs}) from any surface $N=$ constant
 is given by $S=(A_\perp/4G)$ where $A_\perp$ is the area in the transverse direction.  So if we consider a box like region, bounded by $N=0,N=L$ with transverse
area $A_\perp$, then the net flux through $N=0$ surface precisely cancels the flux through $N=L$ surface, giving
$S=0$. Inside the box, $E=0$ and hence, both (\ref{sofe}) and (\ref{denstate}) are trivially satisfied, with {\it arbitrary} value for $\beta$. 
(Since the Rindler horizon is non-compact, it is not possible to ``enclose'' it by a compact 
surface  ${\partial\cal V}$. However, if we use (\ref{sofe}) with the contribution to entropy
from any single surface $S=(1/4G)A_\perp,\beta=2\pi/\kappa$, we get $\kappa=4\pi G(E/A_\perp)$
which is just the expression for gravitational acceleration produced by a plane sheet of matter with energy density
$(E/A_\perp)$. This makes physical sense because the accelerated observer will indeed attribute such a surface
energy density as the source of apparent gravitational acceleration. This interpretation has been noted before in different contexts; see \cite{membrane}).

The existence of the horizon plays a vital role in this description.  If the original metric
had $N({\bf x})>0$, then the original coordinates will take unphysical values if we try to solve for $N({\bf x})=0$ and it seems reasonable to restrict the range of the coordinate $N$. Then, there is no horizon and the value of $\beta$ is undetermined and we should treat this merely as a parameter specifying the range of time integration. (This is analogous to the Rindler frame discussed 
above). The result (\ref{sofe}) will however continue to hold with a well defined value for $E$ but arbitrary value for $S$ due to the arbitrariness in the definition of $\beta$.

Finally we speculate on the possibility of defining the entropy and density of states for a time dependent horizon. The transition to a system of coordinates in which
$N$ is a new coordinate is possible even in time dependent geometries, except that --- since $N=N(t,{\bf x})$ originally ---
we will now generate $g_{0\mu}$ terms. For a wide class of such spacetimes, at least a local patch of horizon can be defined by the $N=0$ surface; but no  natural definition of temperature exists since analytic continuation to Euclidean time etc will not work except possibly in the case of slowly varying temperature. As regards the definition of entropy  we can take
\begin{equation}
S=\frac{1}{8\pi G}\int\sqrt{-g}d^4x\nabla_i (a^i+Ku^i)
\end{equation}
since these two terms arise together in Euclidean gravity from the integral over the $2K$ in the boundary surface
\cite{tpapoorva}. If we take,
\begin{eqnarray}
\beta E&=&2\int d^4x\sqrt{-g} \Big[(T_{ab}-\frac{1}{2}Tg_{ab})u^au^b\nonumber\\
&& \qquad +(^3R-16\pi G T_{ab}u^au^b)\Big]
\label{defofso}
\end{eqnarray}
then
Einstein's equations maintain $S=(1/2)\beta E$ in general. Maintaining the relation
(\ref{ehfree}) will require the identification:
\begin{eqnarray}
\label{defofuo}
U&\equiv&\frac{1}{8\pi G}\int d^4x\sqrt{-g}(^3R -8\pi G T_{ab}u^au^b)\\
&=&\frac{1}{16\pi G}\int d^4x\sqrt{-g}(^3R +K_{ab}K^{ab}-K^2)\nonumber
\end{eqnarray}
which is actually the ADM action.
In static spacetimes $^3R =16\pi G T_{ab}u^au^b$ making Eq.(\ref{defofuo}) match with  Eq.(\ref{defofu}).
The physical meanings of these relations are unclear, especially since they refer to the time dependent system
with no precise meaning for $\beta$; only further investigations will show whether these definitions are useful.


\end{document}